\documentclass[%
 reprint,
superscriptaddress,
 amsmath,amssymb,
 aps,pra,
]{revtex4-2}

\usepackage{graphicx}
\usepackage{dcolumn}
\usepackage{bm}
\usepackage{subcaption}
\usepackage{amsmath}
\usepackage{xfrac}
\usepackage{xcolor}
\usepackage{soul}
\graphicspath{ {./figs/} }

\begin{document}

\preprint{APS/123-QED}

\title{Towards an optical fiber-based temporally multiplexed single photon source}

\author{Agustina G. Magnoni}
\email{amagnoni@citedef.gob.ar}

\affiliation{%
 Laboratorio de Óptica Cuántica, DEILAP, UNIDEF (CITEDEF-CONICET), Buenos Aires, Argentina}%
 \affiliation{Departamento de Física, Facultad de Ciencias Exactas y Naturales, UBA, Ciudad de Buenos Aires, Argentina}
 
\author{Laura T. Knoll}%
\affiliation{%
 Laboratorio de Óptica Cuántica, DEILAP, UNIDEF (CITEDEF-CONICET), Buenos Aires, Argentina}%
 \affiliation{Departamento de Física, Facultad de Ciencias Exactas y Naturales, UBA, Ciudad de Buenos Aires, Argentina}
 
\author{Lina Wölcken}%
\affiliation{Departamento de Física, Facultad de Ciencias Exactas y Naturales, UBA, Ciudad de Buenos Aires, Argentina}

\author{Julián Defant}%
\affiliation{Departamento de Física, Facultad de Ciencias Exactas y Naturales, UBA, Ciudad de Buenos Aires, Argentina}

 \author{Julián Morales}%
\affiliation{%
 Laboratorio de Óptica Cuántica, DEILAP, UNIDEF (CITEDEF-CONICET), Buenos Aires, Argentina}%
  \affiliation{Departamento de Física, Facultad de Ciencias Exactas y Naturales, UBA, Ciudad de Buenos Aires, Argentina}
  
 
 \author{Miguel A. Larotonda}%
\email{mlarotonda@citedef.gob.ar}
\affiliation{%
 Laboratorio de Óptica Cuántica, DEILAP, UNIDEF (CITEDEF-CONICET), Buenos Aires, Argentina}%
 \affiliation{Departamento de Física, Facultad de Ciencias Exactas y Naturales, UBA, Ciudad de Buenos Aires, Argentina}




\date{\today}

\begin{abstract}

We demonstrate the feasibility of implementing a photon source with sub-Poissonian emission statistics through temporal multiplexing of a continuous wave heralded photon source in the optical communications wavelength range. We use the time arrival information of a heralding photon to actively modify the delay of the heralded photon in an all-fiber assembly, in order to synchronize the output with with respect to an external clock. Within this synchronized operating regime we show that the addition of a single temporal correcting stage can improve the figure of merit for single-photon emission of a heralded photon source. We obtain a brightness improvement factor of approximately 1.8 and an enhancement of the signal-to-noise ratio, quantified by the coincidence-to-accidental counts ratio. These results, clear the way for integrated optics non-classical photon sources in the optical communication band.

\end{abstract}

\maketitle


\section{\label{sec:intro} Introduction}

Quantum photonics has emerged as one of the most promising areas of research in quantum technology, with application to secure communications, metrology and quantum information processing. Specifically, photonic systems play a transversal role: photons can be used as the base of a quantum computer, even reaching quantum supremacy \cite{knill2001scheme, zhong2020quantum, o2007optical}, and they also represent the only practical option for communication between quantum nodes \cite{monroe2002quantum}, even when such devices may rely on different physical quantum systems for codification and processing. Quantum cryptography \cite{scarani2009security, pirandola2020advances} and metrology \cite{dowling2015quantum, taylor2016quantum, giovanetti2011, moreau2019imaging} also rely entirely on optical set ups. Optical communications standards impose additional conditions on the wavelength, spatial mode and spectral bandwidth of practical quantum photonic states.

As a consequence on-demand, single-photon sources become a valuable asset for quantum optics-based devices since they exhibit sub-Poissonian statistics, a key feature in quantum metrology applications \cite{berchera2019quantum}. Also, such states are of the utmost importance for the security of Quantum Key Distribution protocols, unless decoy state techniques are applied \cite{decoy}. Globally, a great effort is being made to generate and characterize near optimal single-photon sources, where the current approaches can be divided into deterministic and probabilistic sources. 

Deterministic sources are based on single quantum emitters systems, like  fluorescence of single atoms, ions or molecules \cite{kimble1977photon, mandel1979sub, basche1992photon}, and artificial atoms like quantum dots or nitrogen-vacancy centers \cite{aharonovich2016solid, ding2016demand, somaschi2016near, tomm2021bright}. In contrast, probabilistic sources rely almost entirely on second-order ($\chi^{(2)}$) or third order ($\chi^{(3)}$) nonlinear effects such as Spontaneous Parametric Downconversion (SPDC) and Four Wave Mixing (FWM). Even though the deterministic nature of solid emitters seems a key advantage over the inherently random photon generation of probabilistic sources, in practice a number of experimental factors contribute in the blurring of the difference between both strategies. Extraction inefficiency and the necessity of frequency conversion to obtain telecom-wavelength photons mix the initially deterministic phenomenon with random processes, giving as a result a probabilistic source. 

On the other hand, single-photon sources based on SPDC or FWM operate at room temperature and generate photons naturally at wavelengths suitable for propagation in commercial optical fiber links, a crucial attribute for quantum communications. Here, photons are created in pairs (signal and idler) from an original pump photon. In the simplest experimental implementation, one of the photons is used as a herald or announcer of the presence of the second photon, which is the actual output of the source. In this way, empty pulses can be ideally eliminated, although the emission of non-empty states does not occur at well-defined times; it is rather a probabilistic event. Recent developments in this field include photonics integrated devices, achieving heralding efficiencies exceeding 50\% \cite{bock2016highly, montaut2017high}, even reaching 90\% \cite{paesani2020near} within the chip.

Although the heralding process removes the probability of having empty pulses, since the statistics of the pair emission is nearly poissonian, the probability of having pulses with more than one photon is non-negligible. This probability also scales with the pump beam power\cite{christ2012limits}, so ramping up the power to gain brightness is not a convenient option. More sophisticated devices include the presence of multiplexing stages, both temporal \cite{jeffrey2004towards, mower2011efficient, glebov2013deterministic, schmiegelow2014multiplexing, kaneda2015time, rohde2015multiplexed, zhang2017indistinguishable, kaneda2019high} and spatial \cite{collins2013integrated, mazzarella2013asymmetric}, or even combining both strategies \cite{mendoza2016active, adam2014optimization}: the goal is to increase the probability of emitting a single photon while keeping the probability of multi-photon events low. In this way, the rate at which single photons are emitted can be increased, exhibiting sub-Poissonian statistics: a comprehensive study on the development of such sources can be found in \cite{meyer2020single}. 
Temporal multiplexing has the advantage that there is no need to use more than one pair generating source. In this type of multiplexing, a time interval (temporal bin) is defined such that there is a very low probability, which is given by either a Poisson or a thermal distribution depending on the amount of modes, on each bin to obtain a photon. The key to this strategy is that 
a detection of a \emph{single} herald photon can occur within several consecutive temporal bins, indicating that there is a heralded photon in the system; timing information with respect to an external clock is used to actively route the heralded photon on a specific delay line or optical storage device, that synchronizes the photon output with (a sub-multiple of) the external clock. 

In this work, we present experimental results on a proof-of-principle demonstration of a single-photon source based on a SPDC source and time multiplexing, implemented with optical fiber loops and fast integrated optical switches in the standard optical telecommunications wavelength range. A distinctive feature is that it operates with a continuous wave (CW) pump laser, while most of the studied implementations rely on the utilization of a pulsed pump. The experimental set-up is presented in section II, while sections III and IV show the results and discussion. Gated operation is obtained by means of the fast optical switches action.

\section{\label{sec:fuente} Single-photon source}

\begin{figure}[h]
    \centering
    \includegraphics[width=0.95\linewidth]{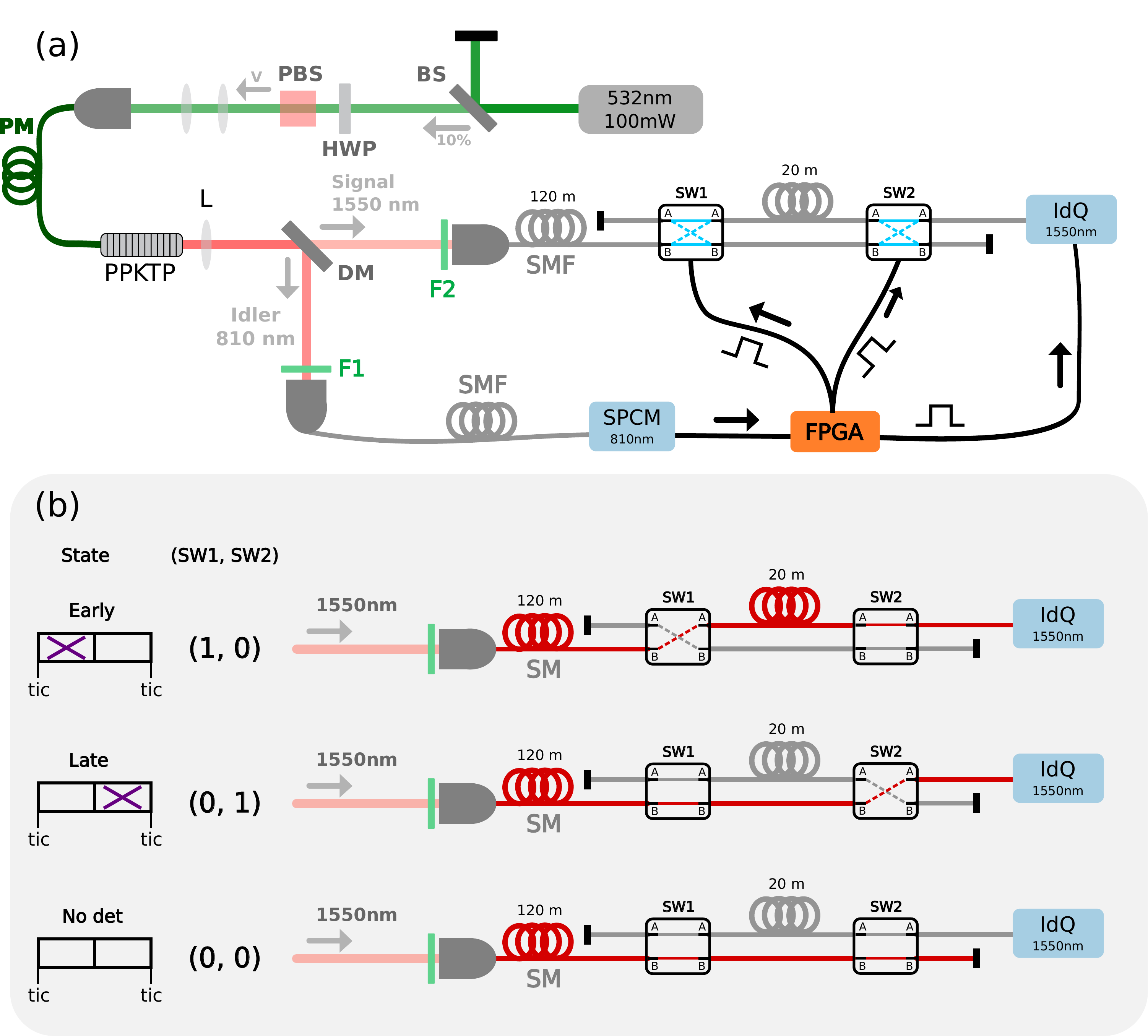}
    \caption{(a) Schematic setup of the single-photon source. Photon pairs at 810~nm (signal photon) and 1550~nm (idler photon) are created by SPDC with a CW green pump. The timing information of the signal photon arrival respect to an external clock is used to configure a time multiplexing system based on two 2$\times$2 optical switches and different available fiber paths. In (b) the principle of operation of the single-stage time multiplexing is depicted. An external clock signal of 5 MHz controls the entire source: if the signal photon is detected in the first temporal window, the long fiber path is enabled; if it is detected on the second one, the idler photons are routed via the (negligibly) short path. Finally, if no signal photon is detected, the idler output remains closed.}
    \label{fig:binmux_specific}
\end{figure}

The sub-poissonian photon source built in this work relies on the idea of achieving single-photon emission from photon pair generation by SPDC and a subsequent time multiplexing stage, based on an array of binary-divided optical delays. This source was proposed and theoretically studied in Ref. \cite{magnoni2019performance}. The idea is to use the arrival information of the herald (signal) photon to impose a specific delay to the heralded (idler) photon, in order to synchronize it with an external clock tick. This delay is defined with an active delay network of optical fiber patchcords. Such process also increases the brightness of the source since it multiplexes the emission of a number of time windows.  In this work, we have experimentally implemented a simplified version that multiplexes single-photons generated in two individual time windows to a common output. A scheme of the experimental set up is shown in figure \ref{fig:binmux_specific} (a).

A periodically-poled, waveguided KTP crystal (PPKTP) is pumped by a 532~nm CW, single longitudinal mode laser (Torus 532, Laser Quantum). The pump power is controlled with a combination of a half wave plate (HWP) and a polarizing beam splitter (PBS). The laser is coupled into the polarization mantaining (PM) pigtail of the crystal. The periodic poling of the crystal allows for collinear quasi-phase match for signal photons at 810~nm and idler photons at 1550~nm.  At the output of the crystal, collinear photon pair beams are collimated with an aspheric lens (L), while the residual pump light is removed and the downconverted photon beams are separated using a dicroic mirror (DM). Each beam is coupled into a single mode fiber (SMF), after passing through spectral filters (F1 and F2) that block unwanted light. Signal photons are filtered with a 10~nm FWHM bandpass filter, while a highpass filter is placed within the idler beam to block residual pump and signal light. It is worth noting that the bandwidth of the generated SPDC photons is not imposed by the filters but by the quasi-phase matching condition in the 25~mm long PPKTP instead. 

Signal photons are directly detected in an avalanche photodiode-based Single Photon Counting Module (SPCM, AQ4C Excelitas) and the arrival information is stored at a field programmable gate array (FPGA) board that controls the time multiplexing circuit and the trigger of the heralded photons detector. Concurrently, idler photons are fed into a 120~m optical delay line to allow for the FPGA to set the switches to the selected fiber path, depending on the heralding detection.

The time multiplexing system is depicted in figure \ref{fig:binmux_specific} (b) and it is based on two 2$\times$2 optical switches (NanoSpeed 2x2 Series Fiber, Agiltron) or commutators: they optically connect two inputs to two outputs, which are exchanged upon a logical electrical TTL input. The minimum width of the individual time multiplexing  window is chosen based on the dynamic properties of these optical switches: given their rise and fall times, we need to use a minimum width of 100~ns. Each optical switch has an insertion loss of around 1 dB and, in order to reduce as much as possible the total loss of the multiplexing stage, each fiber patch is custom made and the connections with the optical switches are fused.  

The principle of operation can be summarized as follows: an external clock signal of period equal to two individual time windows (200~ns) is the time reference. The output of the system only opens during the second time window and the goal is to generate a synchronized output for the idler (heralded) photon whether it is initially present in the first or in the second temporal window. To achieve such behavior the FPGA uses the arrival information of the signal photon to decide whether it is an early or a late photon (detected during the first or the second window). Using this information, the correspondent idler path is enabled by acting on the switches: the 100~ns delay in the ``early'' case or a negligible delay in the ``late'' case. In short, since the system routes the photons present either in the first or in the second window to the output without any pile up, the brightness of the source is increased for the same pump power. This manipulation alters the statistics of the source, turning the original poissonian distribution of an SPDC process with many modes into a sub-poissonian emission \cite{magnoni2019performance}.

Finally, the idler photons exit the source and are detected with a gated SPCM (ID Quantique IDQ 201, IdQ). This device is enabled only at the second time window. Due to the broad bandwidth filtering, the coherence time of the idler photons is much shorter than 100~ns, therefore these photons can be detected anytime during the second time window (we estimate a coherence time of approximately 0.4~ps by measuring a bandwidth of 5~nm of the signal photons and calculating the corresponding idler bandwidth using the energy conservation condition for donwconverted photons, regarding the negligible bandwidth of the single longitudinal mode pump laser).

\section{\label{sec:performance} Performance}

\begin{figure}[h]
    \centering
    \includegraphics[width=0.95\linewidth]{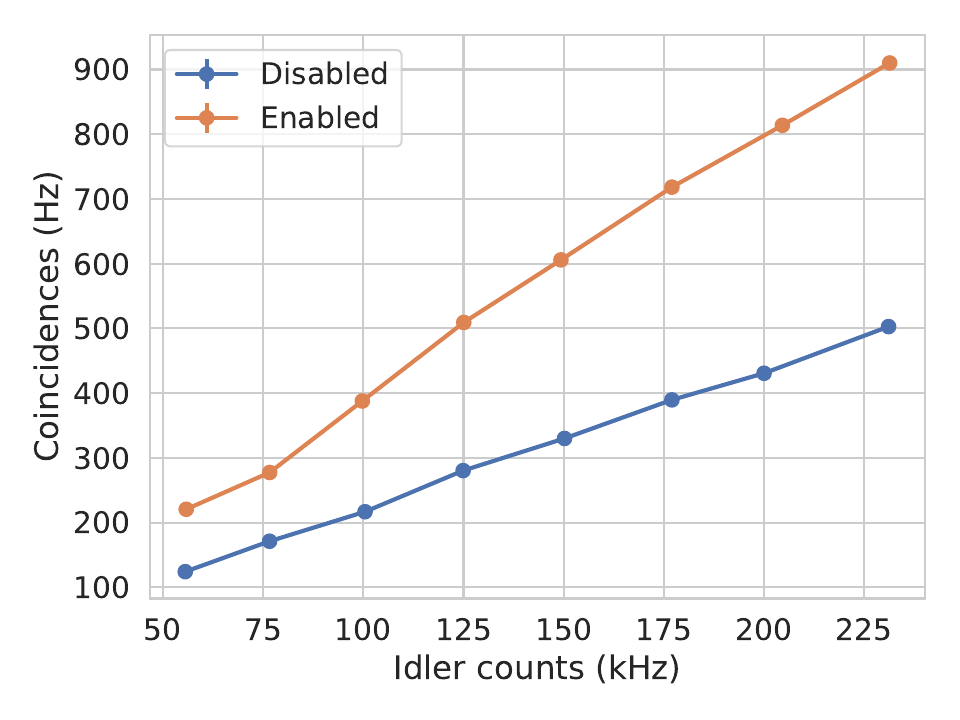}
    \caption{Detected coincidences as a function of the signal counts for the time multiplexing stage enabled and disabled.}
    \label{fig:coincidences_810}
\end{figure}

The inclusion of the multiplexing stage in the single-photon source should produce an enhancement in the output brightness without increasing the multiphotonic component. This results in a modification of the emission statistics. Figure \ref{fig:coincidences_810} shows the measured coincidences as a function of the signal counts, for different pump powers ranging from 0.5~mW to 1.5~mW, with the multiplexing stage enabled and disabled. The coincidences are the detected counts of the 1550~nm photons, triggered by the FPGA electric signal. 

The total amount of light increases linearly with the pump power (and consequently with the signal counts) in both cases. Enabling the multiplexing process generates a significant increase in the 1550~nm detected counts. The number of measured signal and idler counts differs in order of magnitude mainly because of the difference in efficiency, operating principle (gated and free running) and dead time of the detectors, together with the difference in the optical losses of each path.   

\begin{figure}[h]
    \centering
    \includegraphics[width=0.95\linewidth]{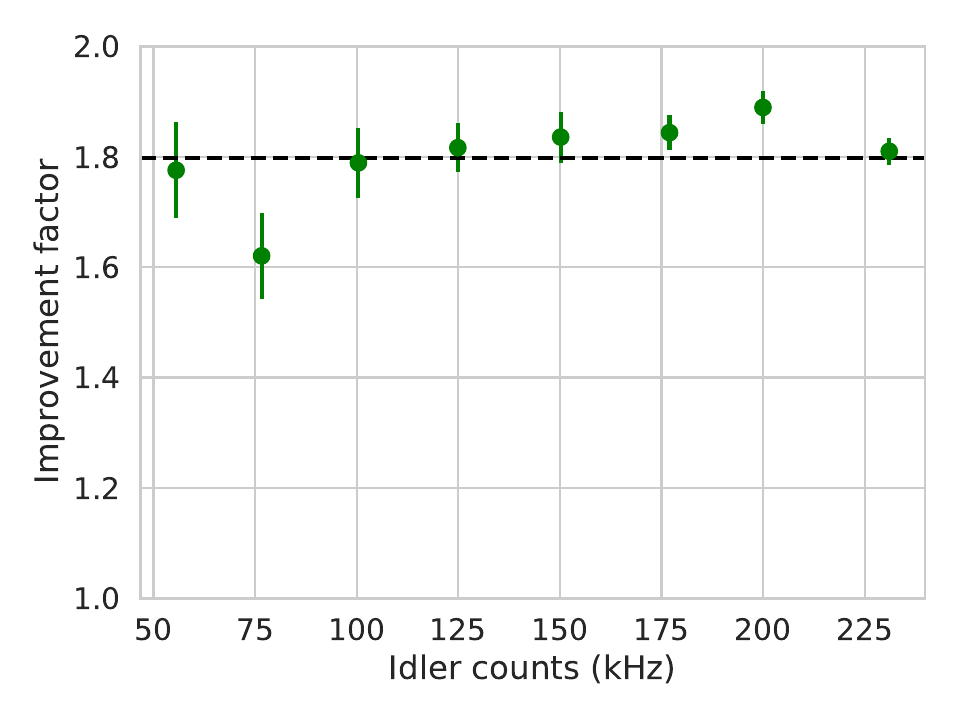}
    \caption{Ratio of coincidences (improvement factor) with enabled and disabled time multiplexing stage as a function of the signal counts. The black dashed line corresponds to the mean improvement factor: $(1.80 \pm 0.03)$.}
    \label{fig:improvement_factor}
\end{figure}

This enhancement can be visualized as the ratio of the detected coincidences or improvement factor for a given amount of signal counts. This is shown in figure \ref{fig:improvement_factor}. This quantity remains constant over the whole range of pump powers studied, with a mean value of $(1.80 \pm 0.03)$. The performance of the time multiplexing stage shows to be nearly optimal since, ideally, the time multiplexing stage should result in a twofold increase.   

In order to do a fair comparison we must consider the same amount of total losses in the idler path. Such conditions can be obtained by simply enabling and disabling the multiplexing system. The comparison with a plain heralding source without the loss inserted by the optical switches can also be done, lowering down the improvement factor to $(1.13 \pm 0.02)$. Nonetheless, our source can eventually be further improved by incorporating a third optical switch to the time multiplexing stage, adding only 1~dB loss but now multiplexing 4 time windows. However, there is a trade off between the amount of correction stages added and the loss inserted, studied in \cite{magnoni2019performance}.  

\begin{figure}[h]
    \centering
    \includegraphics[width=0.95\linewidth]{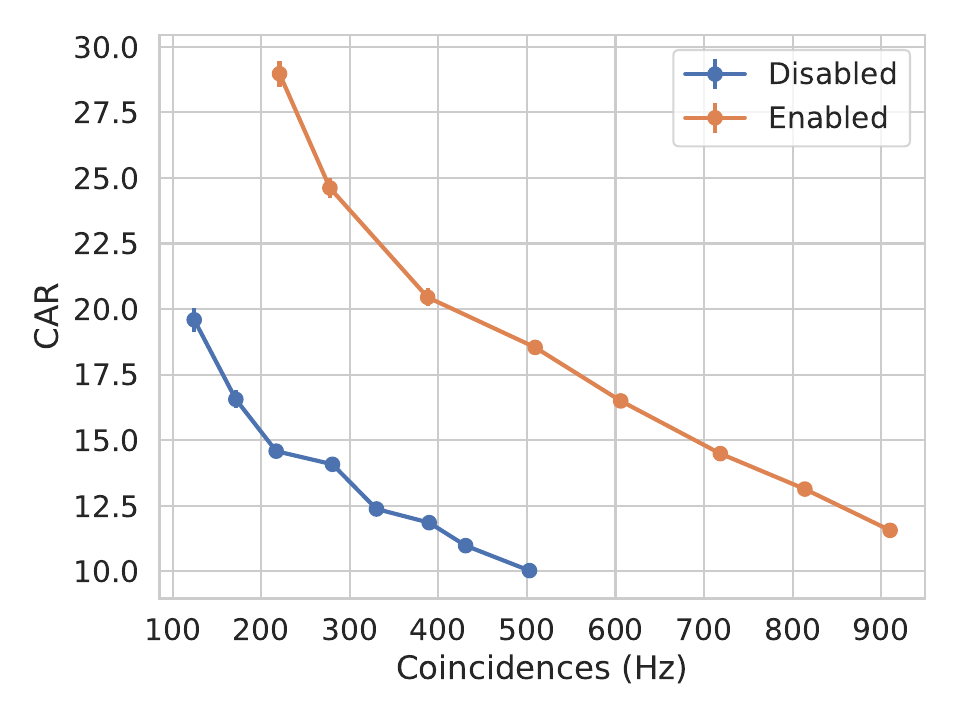}
    \caption{Coincidence-to-accidental ratio (CAR) as a function of coincidence counts, with the time multiplexing stage disabled (red) and enabled (blue). As expected, both curves decrease with the pump power due to the increase of accidental counts. The CAR shows an almost twofold improvement for a given output brightness of the source.}
    \label{fig:CAR_coincidences}
\end{figure}

Regarding the change in the statistics, the coincidence-to-accidental ratio (CAR) is a representative figure of merit of the ratio of heralded photons with respect to the multiphoton emission \cite{xiong2016active}. Accidental coincidences are the detected idler photons that were triggered by an signal photon that is not their correlated pair. Figure \ref{fig:CAR_coincidences} shows this quantity as a function of the coincidences (which can be altered varying the pump power) when enabling and disabling the time multiplexing stage. We thus compare the behavior of the source in both working modes at the same output brightness. 
In both cases the CAR decreases with increased coincidence rate, due to the cuadratic dependence of the accidental coincidences with the single photon counts since this is the case for the original heralded source. However, for a given brightness (coincidence level) the CAR can be increased simply by enabling the time multiplexing stage to nearly twice its value. This indicates a better relation between heralded single-photons and the multi-photon emission. Furthermore, for a target CAR value, the brightness of the source suffers almost a three-fold increase. 

\begin{figure}[h]
    \centering
    \includegraphics[width=0.95\linewidth]{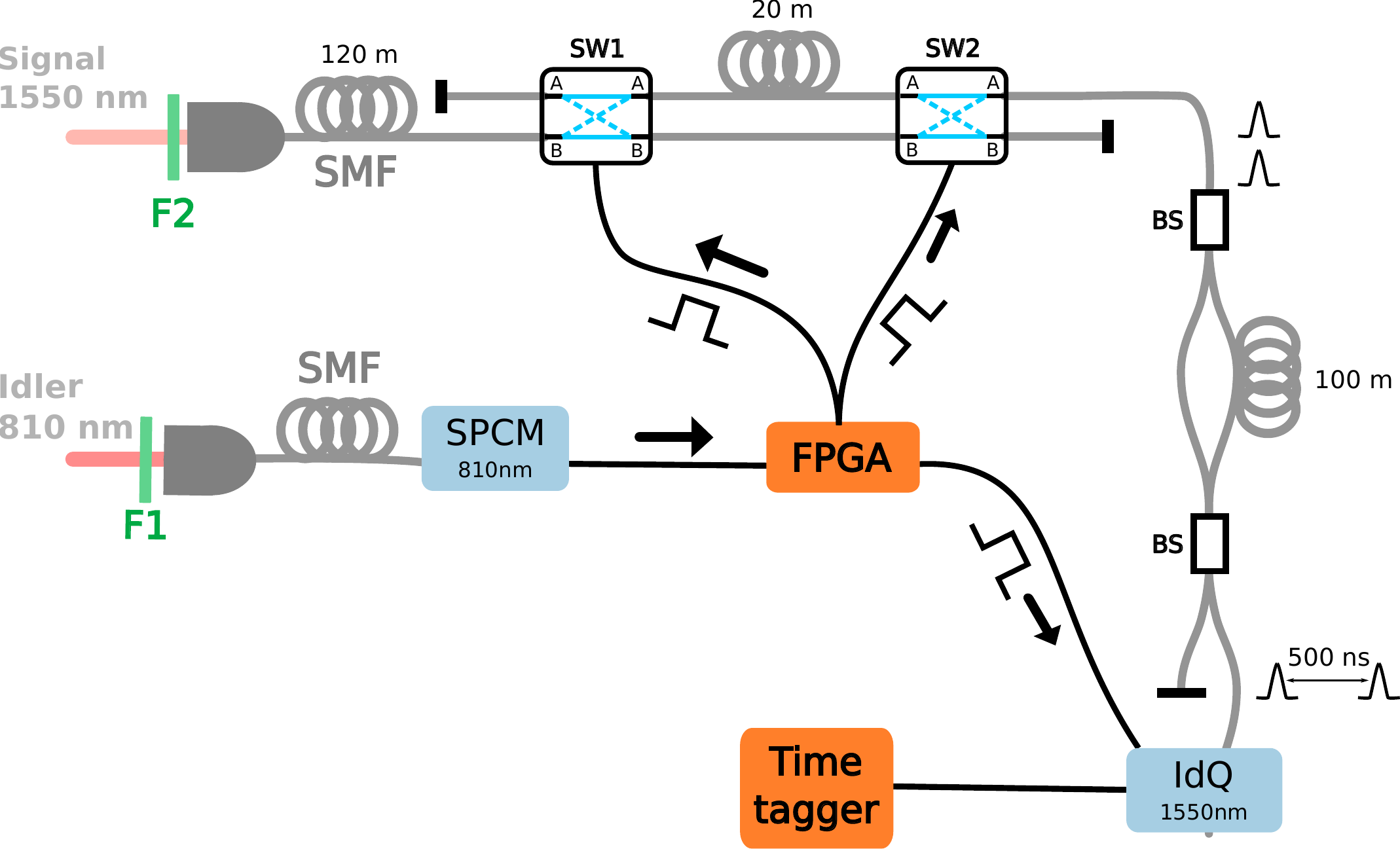}
    \caption{Set up for measuring the second-order coherence function $g^2(0)$ based on a beamsplitter (BS) and a 500~ns delay line, in order to use only one detector (IdQ).}
    \label{fig:setup_g2}
\end{figure}

Multiphoton emission can be quantified by measuring the initial value of the second-order coherence function $g^2(0)$. This measurement can be obtained detecting coincidences when splitting the idler photons output on a beamsplitter, conditioned to the detection of an signal photon. We have implemented the set up shown in figure \ref{fig:setup_g2} to measure the second-order coherence function using only one single-photon detector, employing a time delay strategy with an optical fiber spool of approximately 500~ns total delay.   

The second-order coherence function can be expressed in terms of the coincidences between the signal and each one of the idler ports $(R_{12} \text{ and } R_{13})$, and the triple $(C_{123})$ coincidences (between the signal and the two outputs of idler photons). Also, the heralding signal counts $H$ has to be taken into consideration: 

\begin{equation}
\label{ec:g2}
    g^2(0) = \frac{C_{123} H}{R_{12} R_{13}}.
\end{equation}

\begin{figure}[h]
    \centering
    \includegraphics[width=0.95\linewidth]{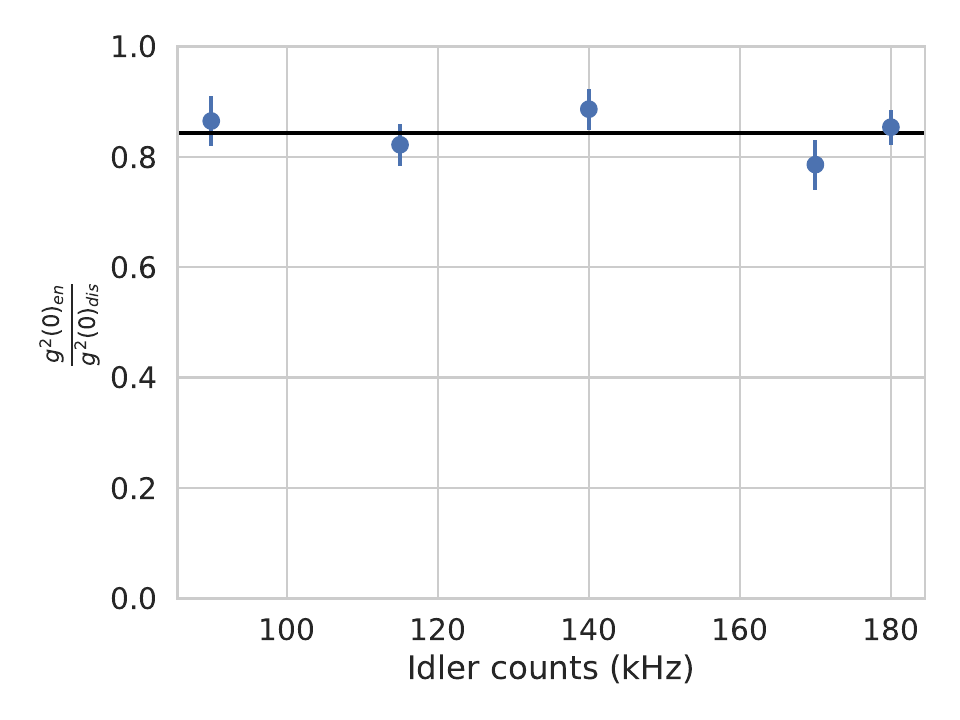}
    \caption{Ratio of the second-order coherence function $g^2(0)$ when enabling and disabling the time multiplexing stage as a function of the signal counts. The solid line represents the mean of the data.}
    \label{fig:g2}
\end{figure}

This expression needs an additional correction that accounts for the different modes of operation of the two detectors (gated for the idler photons and free running for the signal photons), and also for their differences in deadtimes. Rather than estimating these corrections, we can obtain the ratio of $g^2(0)$ in the two modes of operation: as a single heralded source and as a multiplexed (heralded) source. In this way, corrections due to detectors are cancelled since the same experimental setup is used in both regimes.   Figure \ref{fig:g2} shows the ratio of the second-order coherence function with the multiplexing stage enabled and disabled as a function of the signal counts: the increase in brightness due to the addition of the multiplexing scheme does not entail an augmented multi-photon emission. 

Even though the brightness is improved approximately a factor 1.8 (see figure \ref{fig:improvement_factor}) the $g^2(0)$ does not suffer an increase when enabling the time multiplexing stage and even has a small diminish. This is consistent with both that the multi-photon emission remains constant (since it depends on the pump power for a fixed temporal window), and that the zero-photon component is indeed reduced (since the temporal window for detection is doubled in length) to a level corresponding to a single temporal window. 
\section{\label{sec:final} Final remarks and discussion}

We have implemented a time multiplexing strategy on a telecom CW heralded single-photon source in order to increase the brightness without increasing the multi-photon emission contribution. The multiplexing stage is built on fiber optics and integrated optical commutators which makes it robust and scalable and , since free-space optics are not the dominant part of the set-up. Further integration may be attainable by fiber coupling the output of the SPDC source and using wavelength division demultiplexers to separate the idler and signal beams.

We studied the behavior of the source with one correction stage, thus multiplexing the emission of two single temporal windows, obtaining an improvement in the brightness of the source of around 1.8 (figure \ref{fig:improvement_factor}) that is nearly optimal. The signal-to-noise ratio is also enhanced, since the relation between the heralded and the accidental counts increases. This is shown in figure \ref{fig:CAR_coincidences}, that illustrates how the coincidence-to-accidental ratio nearly doubles for the same output brightness (given by the coincidence rate). Finally, the second-order degree of coherence is compared with the multiplexing stage enabled and disabled, showing a noticeable reduction of the $g^2(0)$ value. 

These results correspond to a first step towards the development of a telecom-wavelength single-photon source, based principally on fiber integrated components, easy to operate and potentially scalable. We show that the multiplexing scheme shows enhancement even when the source is operated with a CW pump, an infrequent feature in this kind of source since they usually rely on a pulsed pump with the period acting as the time reference. A pulsed pump with a pulse duration shorter than the coherence time of the downconverted photons will further improve the behavior of the scheme as a single photon source. The implementation of a second correcting stage, which will allow to multiplex four temporal windows is currently in progress. 

\bibliography{binmux_biblio}

\end{document}